\documentclass[12pt]{article}
\usepackage{graphicx,psfrag,epsfig, color}
\usepackage{amssymb,amsmath,amscd,amsthm}
\usepackage{graphicx,psfrag,epsfig}
\usepackage{hyperref}

\date{}

\begin{document} \author{ Dmitri Goloubentsev\thanks{MathLogic LTD, London, UK., dmitri@matlogica.com}, Evgeny Lakshtanov\thanks{	CIDMA, Department of Mathematics, University of Aveiro, Aveiro 3810-193, Portugal and MathLogic LTD, London, UK., lakshtanov@matlogica.com}}

\title{Remarks on Automatic Adjoint Differentiation for gradient descent and models calibration}

\maketitle

\begin{abstract}
 In this work, we discuss the Automatic Adjoint Differentiation (AAD) for functions of the form $G=\frac{1}{2}\sum_1^m (Ey_i-C_i)^2$, which often appear in the calibration of stochastic models. { We demonstrate that it allows a perfect SIMD\footnote{Single Input Multiple Data} parallelization and  provide its relative computational  cost. In addition we demonstrate that this theoretical result is in concordance with numeric experiments.}
\end{abstract}

\vspace{0.5cm}

%\section{Introduction}
The Automatic Adjoint Differentiation (AAD) is a rapidly growing field with a wide range of applications including image restoration \cite{goss}, computer vision \cite{Pock} and maching learning in general see e.g. \cite{baydin},  \cite{Kukelova}. The long list of AAD applications can be found in the site of {\it Autodiff} community, see \cite{Autod}.

For the additional reading one can mention monograph by Antoine Savine \cite{savine}, articles by L.Capriotti \cite{capr} and by other authors e.g. \cite{savine2},~\cite{savine3}.

The AAD has become a widespread tool in applications due to the following property:  
If one has an algorithm for a function $f: \mathbb R^n_x \to \mathbb R^m_y$, then AAD maps each $\lambda \in \mathbb R^m$ to the linear combinations of $\partial_i f$
\begin{equation}\label{0711A}
\left \{\lambda \cdot \frac{\partial f(x)}{\partial x_i}, \quad i=1\ldots n\right \}    
\end{equation}
the computation cost does not exceed that of $f$ which is usually between 2 and 10 (depending on a specific AAD tool used). 

The utilization of parallel computations appears naturally in calculation of expectations since it assumes the numerous independent calculation of an integrand. Suppose that one needs to calculate the $\frac d {dx} Ey$, the algorithm for the $\frac d {dx}y$ is first determined and then the average using parallel computations. From that point of view, the AD should avoid differentiate expectations. However, in case of the Adjoint AD, this approach becomes mandatory.  The point is that the adjoint differentiation algorithm is not local (e.g. \cite{capr}) i.e. it requires analysis of the whole algorithm for $f$, thus the differentiation of expectation requires much more memory than taking expectation of the derivative. 

Some important problems involve computations that include an expectation as an intermediate operation.  It is not a trivial question how one can use AAD and avoid differentiation of expectations. In a recent article \cite{Friez}, Fries suggest a general recipe for this problem (see Appendix). However, Fries did not provide the accurate analysis for the computational cost of the proposed algorithm. This work conducts a step-by-step computation of the following simple but important example. 

Consider a functional of the type 
\begin{equation}\label{241118A}
G=\frac{1}{2}\sum_1^m (Ey_i-C_i)^2,
\end{equation}
where $y_i$ are random variables on a filtered probability space $(\Omega,\mathbb Q,\{\mathcal F_t\})$ and $C_i$ are some given target conditions, $i =1,\ldots n$. It is assumed that we are provided with a {\it forward } algorithm 
$$
F ~:~ \mathbb R^{M+N} \to \mathbb R^m_y,
$$  
which calculates $y_i$ for a given set of $M$ parameters and $N$ independent random variables in frames of a time-discretization of the stochastic process. 

We assume that the AAD tool provides the algorithm $R$ (so called "reverse" algorithm, e.g. \cite{capr}) which calculates the full set of derivatives $(\ref{0711A})$\footnote{it can be applied only after the $F$ was launched.}  We also assume that the AAD tool produces parallelized  versions $F_v$ and $R_v$. This means that 
$F_v$ (or $R_v$) provides the execution of $F$ (or $R$) for $c$ independent sets of input data. 
For example, the natural value of $c$ for the case of an AAD-tool tuned to the Intel AVX512 architecture is  
\begin{equation}\label{1511A}
c=8 \times \rm{Number\_of\_Cores}. 
\end{equation}
Is noted that the first factor in (\ref{1511A}) equals to $1$ in all known to us C++ AAD tools. 
%In some cases, the C++ compiler can take advantage of SIMD vectorization and makes this factor greater than one.

We consider the operations $F \to F_v$ and $F \to R_v$ as a simple way to parallelize calculations and we are interested to check if there is a way to use this tool effectively. 

Correction coefficients $K_F$ and $K_R$ are defined via the following  expression 
\begin{equation}\label{1311A}
\rm{Cost(F_v)}=\frac{K_F}{c}{\rm{Cost(F)}}, \quad 
\rm{Cost(R_v)}=\frac{K_R}{c}{\rm{Cost(F)}}
\end{equation}
and they reflect the quality of the $AAD$ tool. Ideally, $K_F=1$, but in practice, the different software optimization and different hardware specifics can never make it perfect. One reason is that the  executable version of $F$ is produced with all compiler's optimization abilities turned On, including the ability SIMD vectorization.  

{\bf Some remarks on the Monte-Carlo simulations}. We approximate expectations by 
$$
Ey\sim \frac{1}{\rm{number\_of\_Paths}} \sum_{k=1}^{\rm{number\_of\_Paths}}  y(\omega_k),
$$
where the set $\{\omega_i\}$ contains Monte-Carlo simulations of a given stochastic process and each $\omega_i$ is a simulated sample path of sequences of random variables. It is assumed that drawings are $\mathbb Q$-uniform.  

In particular,the cost of the independent evaluation for the set $\{Ey_i, i=1,\ldots,m\}$ is 
\begin{equation}\label{1211A}
\rm{Cost(\{Ey\})} = \frac{K_F}{c} \times  \rm{number\_of\_Paths} \times\rm{Cost(F)}.  
\end{equation}

{\bf Calculation of the gradient of $G$ (introduced in (\ref{241118A})).} For any parameter $x_j$  we get that
$$
\frac{\partial G}{\partial x_j} = E\left (\sum_{i=1}^m 
(Ey_i-C_i) \frac{\partial y_i}{\partial x_j} \right ),  \quad j=1, \ldots,M.
$$
It leads to the following algorithm :

\begin{enumerate}
\item Calculate $Ey_i$ using $F_v$. The costs of this calculation is given by (\ref{1211A}). 

\item Fix a path $\omega$. Using formula (\ref{0711A}) for the vector $\lambda$ with components $\lambda_i=Ey_i-C_i$ we get the set 
$$
\left \{\sum_{i=1}^m 
(Ey_i-C_i) \frac{\partial y_i(\omega)}{\partial x_j}, \quad j=1\ldots M \right \}   
$$
For $c$ paths, it costs $(K_F+K_R)\rm{Cost}(F)$ since for each path $\omega$ the reverse algorithm can be executed only after the forward algorithm has been launched, unless execution results of the first step can be stored in the memory for each $\omega$. In case memory usage is not constrained, the cost is $K_R\rm{Cost}(F)$
 
\item  Summation over paths $\omega$.  To calculate the cost of this operation we take into account that the integrand should be calculated $\rm{number\_of\_Paths}$ times. 
 
\end{enumerate}

Summarizing, we get the total cost 
\begin{equation}\label{0711C}
\rm{Cost(G)} = \frac{2K_F+K_R}{c} \times \rm{number\_of\_Paths} \times
 \rm{Cost}(F). 
\end{equation}
For C++, AAD-compiler produced by MathLogic LTD can be rewritten as 
$$
\rm{Cost(G)} = \frac{2K_F+K_R}{8\times \rm{number\_of\_Cores}} \times \rm{number\_of\_Paths} \times
 \rm{Cost}(F). 
$$
where $8$ reflects that the AVX512 architecture allows $8$ doubles per vector register.

{\bf Test on the Heston Stochastic Local Volatility  model calibration.} 
For financial institutions, the model calibration is a routine day-to-day procedure. Although the Partial Differential Equation (PDE)-based techniques  became quite popular e.g. \cite{PDE1}, \cite{PDE2}, practitioners more often use direct numeric simulations due to the simplicity of the technique. 

Consider the Heston SLV model given by a process
$$
\left \{
\begin{array}{l}
dS_t= \mu S_tdt+ \sqrt{V_t} L(t,S_t)S_tdW_t^S, \\
dV_t=\kappa(\theta-V_t)dt + \xi \sqrt{V_t} dW_t^V, \\
dW_t^S dW_t^v=\rho dt.
\end{array}
\right .
$$
Our implementation utilizes the standard Euler discretization (e.g. 3.4 of \cite{Glass}) and European options in the quality of the $y_i, i=,\ldots,m$. The set of optimization parameters consists values of the piecewise constant Leverage function $\{L(t_i,S_j)\}$ (where the set of interpolation nodes $\{t_i,S_j\}$ is fixed) and five standard Heston model parameters. We applied the AADC\footnote{C++ AAD-tool produced by matlogica.com} by MathLogic LTD and observed the following values for coefficients $K_F$ and $K_R$  defined in (\ref{1311A}). 
$$
\begin{tabular}{|l|c|c|c|}
\hline
N.Cores=1 & $K_F/c$ & $K_R/c$ & $\frac{\rm{Cost(G)}}{\rm{number\_of\_Paths} \times
 \rm{Cost}(F)} $\\
\hline
AVX2 & 0.39 &0.23 & 1.01\\
\hline
AVX512 & 0.23 & 0.12& 0.58 \\
\hline
\end{tabular}
$$
Tests were executed on one core. The values of the aforementioned coefficients became almost constant when number of time intervals and number of optimization instruments $m$ grow.  As mentioned, all optimization parameters (like vectorization) were turned on while the evaluation of the $\rm{Cost(F)}$.

{
{\bf Conclusion.} We consider a situation when a straighforward application of the AAD software is not possible  i.e. for functions of the form $G=\frac{1}{2}\sum_1^m (Ey_i-C_i)^2$. We propose a two-step algorithm which leads to the estimate (\ref{0711C}) and therefore, provides a perfect SIMD parallelization. A numeric implementation of the algorithm is realized for the Heston model using the AAD-Compiler by matlogica.com. Presented results are in good concordace with (\ref{0711C}). 
}
\hspace{0.5cm}

{\bf Appendix.} Expected Backward Automatic Differentiation Algorithm  following \cite{Friez}.
Consider a scalar function $y$ given by a sequence of operations:
\begin{eqnarray}
y:=x_N \quad ~ \quad \quad \quad \quad \quad \quad \quad \quad \quad \quad \quad \quad \quad \\
x_m:= f_m(x_{\tau_m 1}, \ldots, x_{\tau_m i_m}), \quad 1 \leq m <N. 
\end{eqnarray}
where the number of variables $i_m$ is either 0 and it means that the $x_m$ is an independent variable or $1 \leq i_m < m$. Evidently  for function $\tau_m$ we have $1\leq \tau_m  <m$.  We assume that $k-$th operator is an expectation operator and others $f_m, m\neq k$ given by a closed-form\footnote{i.e. it can be evaluated in a finite number of the "well-known" operations, see e.g. wikipedia https:\slash \slash en.wikipedia.org\slash wiki\slash Closed-form\_expression for the detailed definition.}.

Sequentially,
\begin{itemize}
\item 
Initialise $\overline D_N=1$ and $\overline D_m=0, ~ m<N$.
\item 
For all $m = N, N -1, . . . , 1$ (iterating backward through the operator list)
\begin{itemize} 
\item for all $j = \tau_m 1, . . . , \tau_m i_m $ (iterating through the argument list)
$$
\overline{D}_j =\left \{ 
\begin{array}{ll}
\overline{D}_j+\overline{D}_m \frac{\partial f_m}{\partial x_j}, & m \neq k,\\ 
\overline{D}_j + E\overline{D}_m, & m=k.
\end{array} \right .
$$
\end{itemize}
\end{itemize}
Then, for all $1 \leq i \leq N$ 
$$
E \frac{\partial y}{\partial x_i} = E  \overline{D}_i.
$$

{\bf Acknowledgments.}
 E.L. was partially supported by Portuguese funds through the CIDMA - Center for Research and Development in Mathematics and Applications and the Portuguese Foundation for Science and Technology (``FCT--Fund\c{c}\~{a}o para a Ci\^{e}ncia e a Tecnologia''), within project UID/MAT/ 0416/2019.

\end{document}